\documentclass[12pt]{iopart}
\usepackage{iopams,setstack,graphicx,multirow,hyperref,doi}

\hypersetup{
    colorlinks=true,       
    linkcolor=blue,          
    citecolor=blue,        
    filecolor=blue,      
    urlcolor=blue           
}

\graphicspath{{./}{./figs/}}

\begin{document}
\title[Miniaturizing cold atom technology for deployable vacuum metrology]
{Challenges to miniaturizing cold atom technology for deployable vacuum metrology}
\author{Stephen Eckel, Daniel S. Barker, James A. Fedchak, Nikolai N. Klimov,
Eric Norrgard, Julia Scherschligt}
\address{Sensor Sciences Division, National Institute of Standards and
Technology, Gaithersburg, MD 20899, USA}
\author{Constantinos Makrides, Eite Tiesinga}
\address{Joint Quantum Institute, National Institute of Standards and
Technology and University of Maryland, Gaithersburg, MD 20899, USA}
\ead{stephen.eckel@nist.gov}
\begin{abstract}
Cold atoms are excellent metrological tools; they currently realize SI time and,
soon, SI pressure in the ultra-high (UHV) and extreme high vacuum (XHV) regimes.
The development of primary, vacuum metrology based on cold atoms currently falls
under the purview of national metrology institutes.   Under the emerging
paradigm of the ``quantum-SI'', these technologies become deployable (relatively
easy-to-use sensors that integrate with other vacuum chambers), providing a
primary realization of the pascal in the UHV and XHV for the end-user.  Here, we
discuss the challenges that this goal presents.  We investigate, for two
different modes of operation, the expected corrections to the ideal cold-atom
vacuum gauge and estimate the associated uncertainties.  Finally, we discuss the
appropriate choice of sensor atom, the light Li atom rather than the heavier
Rb.
\end{abstract}
\submitto{\MET}
\maketitle

\section{Introduction}
The emerging paradigm of the Quantum-SI focuses on building devices that obey
three basic ``laws'':  (1) the sensor must be primary, (2) the sensor must
report the correct quantity or no quantity at all, and (3) the uncertainties
must be quantified and fit for purpose.  Cold atoms represent a useful tool in
developing Quantum-SI-based devices because they can be exquisitely manipulated
and controlled.  Deployable cold-atom sensors have the potential to
revolutionize many types of Quantum-SI based measurements such as time, inertial
navigation, and magnetometry.  Here, we focus on the difficulties of
miniaturization of cold-atom technologies for the purposes of vacuum metrology
in the ultra-high vacuum (UHV, $p<10^{-6}$~Pa) to extreme high vacuum (XHV,
$p<10^{-10}$~Pa) regimes.

A cold-atom vacuum gauge is based on the observation that the main source of
atom loss from a cold-atom trap is collisions with background
gas~\cite{Migdall1985,Bjorkholm1988,Willems1995,OHara1999,Arpornthip2012,Yuan2013,Booth2014,Moore2015,Makhalov2016}.  Because cold-atom
traps tend to be shallow ($W/k_B\lesssim 1$~K, where $W$ is the trap depth and
$k_B$ is Boltzmann's constant) compared to room temperature, the vast majority
of such collisions cause ejection of cold atoms from the trap.  This random loss
is well-characterized by an exponential decay of the trapped atom number with
time.  We are currently developing a laboratory-based cold-atom vacuum standard
(CAVS) that will represent a primary standard for the pascal in the UHV and XHV
ranges.  This device will be capable of cooling and trapping different sensor
atoms, including $^6$Li, $^7$Li, $^{85}$Rb, and $^{87}$Rb.

The dominant background gas in vacuum chambers operating in the UHV and XHV
regimes is H$_2$.  The determination of the loss rate coefficient for
$^6$Li+H$_2$ is, in principle, a tractable calculation, and therefore
establishes the primary nature of the CAVS.  Extension to other background and
process gases and to other sensor atoms will be accomplished by measurement of
relative gas sensitivity coefficients (ratios of loss rate
coefficients)~\cite{Scherschligt2017}.

The laboratory-scale CAVS currently in development at NIST is not deployable; it
is neither portable, small, nor easy to use.  It currently occupies an optical
table with roughly 2~m$^2$ of area.  A large experiment is required because of
the large number of components needed to laser cool and trap atoms.  First,
atoms can only be trapped in UHV environments, generally requiring a large
vacuum chamber with ion or getter pumps.  Second, the workhorse of laser
cooling, the three-dimensional magneto-optical trap (3D-MOT), requires optical
access from six directions along three spatial axes.  Third, generally good
magnetic field stability is required, typically obtained by using large coils
that cancel local magnetic fields and gradients.  Shrinking the CAVS to
something deployable thus represents an impressive challenge.  Despite the
difficulties, mobile cold atom systems have been constructed (e.g., an
atom-based accelerometer~\cite{Hauth2013}), and miniaturization continues to be
an active area of research (for example, a proposal to construct a fully
integrated chip-scale device~\cite{Rushton2014}).

Presently, the most-widely-used gauge in the UHV and XHV regimes is the
non-primary Bayard-Alpert ionization
gauge~\cite{Bayard1950,Redhead1966,Arnold1994}, which requires $30$~cm$^3$ and
is controlled using a 2-U standard size rack-mountable controller.  Thus, to
make a deployable, cold-atom based gauge, we tailor our design to occupy a
similar vacuum footprint\footnote{We focus our efforts on the development of
traps and in-vacuum components, rather than on miniaturizing laser systems and
associated electronics.  In general, commercial rack-mountable laser systems
already exist.}.

\begin{figure}
	\center
	\includegraphics[width=0.75\textwidth]{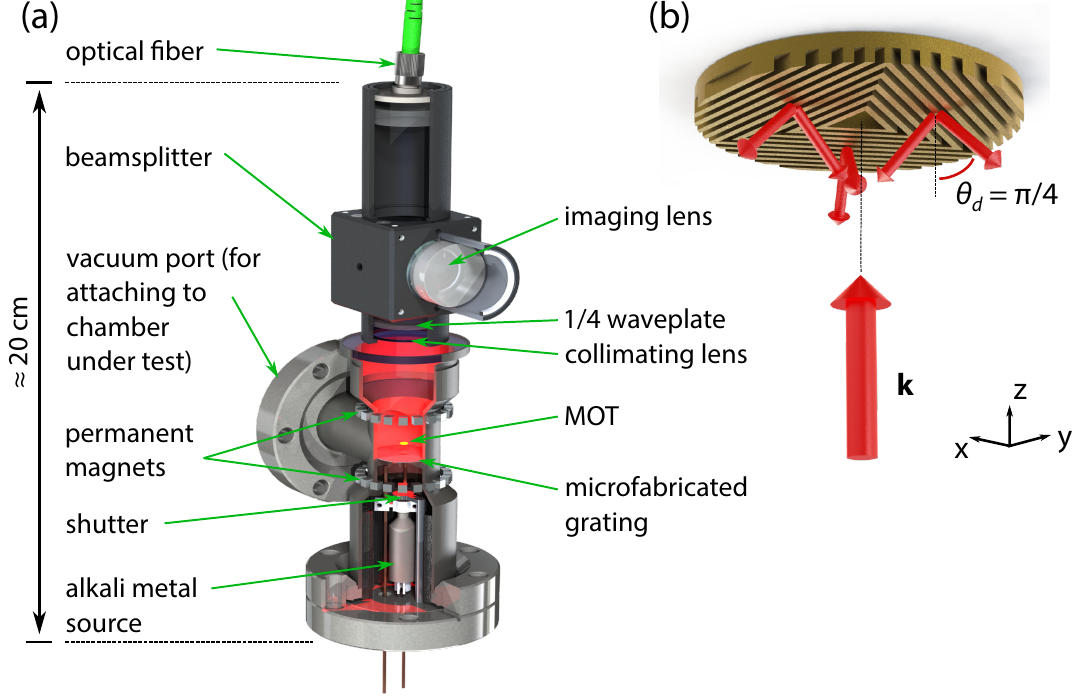}
    \caption{\label{fig:basic_idea} (a) Concept of the p-CAVS, a cold atom-based
vacuum sensor that has the same vacuum footprint as a typical ion or extractor
gauge. (b) Geometry of our grating MOT. A single laser beam (large, red arrow)
traveling along $\hat{z}$ is diffracted into six different beams (small, red
arrows) by three reflective, gold diffraction gratings whose lines form
superimposed triangles and diffract light at $\theta_d = \pi/4$ with respect to
the normal of the grating ($-\hat{z}$). The lines of the diffraction grating are
not to scale.}
\end{figure}

Our current design for a portable CAVS (herein referred to as p-CAVS),  shown in
Fig.~\ref{fig:basic_idea}, is under active development.  Currently, many of its
individual components are being tested separately, and, as such, the final
design is still in flux.   At its core, it uses a micro-fabricated diffraction
grating that generates the necessary spatial beams for laser cooling and
trapping~\cite{Lee2013,Nshii2013}.  This planar MOT is a variant of previously
developed non-planar MOTs like tetrahedral~\cite{Vangeleyn2009} and pyramidal
MOTs~\cite{Lee1996}.  The p-CAVS can create both a magneto-optical trap and a
quadrupole magnetic trap, yielding two possible modes of operation.  In this
paper, we focus on the physical principles for its operation and the associated
uncertainties (Sec.~\ref{sec:princ_of_operation}).   Secondly, we describe some
of the technical design features and their motivation.  These choices depend on
the requirements for a {\it deployable} vacuum gauge, including how it will be
used and treated in the field (Sec.~\ref{sec:field_operation}).  We conclude by
motivating our choice of atomic species (Sec.~\ref{sec:conclusion}).  We include
a short appendix describing the atomic physics used within this paper.
Throughout the paper, we focus primarily on type-B uncertainties and assume
$k=1$.  Type-A uncertainties are briefly discussed in
Sec.~\ref{sec:princ_of_operation_uncertainties}.

\section{Principle of operation and associated uncertainties}
\label{sec:princ_of_operation}
The number of cold atoms $N(t)$ in a trap decays exponentially due to collisions
with background gas molecules, i.e. $N(t) = N_0 e^{-\Gamma t}$, where $\Gamma =
n \langle K \rangle$ is the loss rate, $K=v\sigma$ is the loss rate coefficient,
$n$ is the number density of the background gas, $\sigma(E)$ is the total cross
section for a relative collision energy $E=\mu v^2/2$ and relative velocity $v$.
Here, $\mu$ is the reduced mass, $N_0$ is the initial number of trapped cold
atoms, and $\langle\cdots\rangle$ represents thermal averaging.  In the XHV and
UHV regimes, the ideal gas law is an excellent equation of state of the
background gas, and thus we can relate the loss rate to the pressure through
\begin{equation}
	\label{eq:pressure}
	p = \frac{\Gamma}{\left<K\right>} k_B T,
\end{equation}
where $T$ is the temperature of the background gas.  Equation~\ref{eq:pressure}
represents the ideal operation of the CAVS and p-CAVS.

Perhaps the most crucial quantity in Eq.~\ref{eq:pressure} is $\left<K\right>$.
We described the techniques for determining this quantity in a previous
work~\cite{Scherschligt2017}.  We intend to calculate {\it a priori} the
collision cross section for $^6$Li+H$_2$.  For other gases, we plan to measure
the ratio of loss rate coefficients to that of $^6$Li+H$_2$.  In the present
work, we will assume the uncertainty in $\left<K\right>$ to be $5$~\%, an
estimate based on the expected results of a laboratory-scale CAVS.  Both
theoretical scattering calculations and experimental work are ongoing.

\begin{table}
	\center
	\begin{tabular}{l|r|r|r|r|}
		& Li (2S) & Li$^*$ (2P) & Rb (5S) & Rb$^*$ (5P) \\
	\hline
	H$_2$~\cite{Zhu2002} & 83 & & 160 & \\
	He~\cite{Jiang2015,Tao2012} & 23 & & 45 & \\
	H$_2$O & 150 & 100 & 280 & 280 \\
	N$_2$ & 180 & 130 & 350 & 350 \\
	O$_2$ & 160 & 120 & 310 & 310 \\
	Ar~\cite{Jiang2015,Tao2012} & 180 & & 340 & \\
	CO$_2$ & 270 & 190 & 520 & 510 \\
	\hline\hline
	\end{tabular}
	\caption{\label{tab:polarizability} Estimated $C_6$ coefficients in atomic
units.  Entries without references were calculated using the Casimir-Polder
integral, for which we estimate a 10~\% uncertainty for the values.  The
coefficients do not depend on isotope to the accuracy given.}
\end{table}

{\it Ab initio} quantum-mechanical scattering calculations are difficult, but we
can estimate the cross section using semiclassical
theory~\cite{child2014molecular,landau2013quantum} for a cold, sensor atom of
mass $m_c$ and a (relatively-hot) room-temperature background-gas atom or
molecule of mass $m_h$.  In this theory, the isotropic, long-range attractive
part of the inter-molecular potential fully determines the total elastic cross
section.  This part of the potential is dominated by a van der Waals interaction
$-C_6/r^6$, where $C_6$ is the dispersion coefficient and $r$ is the separation
between the cold atom and the background gas molecule.
Table~\ref{tab:polarizability} lists $C_6$ for various combinations of cold
atoms (both ground S and first excited P states) and background gases as
calculated using the Casimir-Polder relationship,
\begin{equation}
	C_6 = \frac{3}{\pi}\int_0^\infty \alpha_A(i\omega)\alpha_B(i\omega)\ d\omega
\end{equation}
for species $A$ and $B$.  Accurate dynamic polarizabilities $\alpha(\omega)$ as
a function of frequency $\omega$ exist for each alkali atoms' ground
state~\cite{Derevianko2010}.  The dynamic polarizability of the excited state
has been calculated for Li (2P$_{3/2}$)~\cite{Tang2010} and can be inferred from
transition frequencies and matrix elements for Rb
(5P$_{3/2}$)~\cite{Safronova2011}.  For common background gases, we use dynamic
polarizabilities found in the literature for water~\cite{Mata2009},
nitrogen~\cite{Oddershede1982}, oxygen~\cite{Hohm1994}, and carbon
dioxide~\cite{Hohm1994}.  For Li, the dispersion coefficient is a factor of two
smaller than Rb for the same background molecule.  Coincidentally, there appears
to be little to no difference in the $C_6$ coefficients for the 2P and 2S states
of Rb.

Within the semiclassical theory~\cite{child2014molecular,landau2013quantum}, we
calculate both the differential and total cross sections from the semiclassical
phase shift for partial wave $\ell$,
\begin{equation}
	\label{eq:phase_shift}
	\eta_\ell(E) = \frac{32\pi}{3}\frac{(E/E_6)^2}{\ell^5},
\end{equation}
where $E_6 = \hbar^2/(2\mu x_6^2)$ is the van der Waals energy, $x_6 = (2\mu
C_6/\hbar^2)^{1/4}$ is the van der Waals length, and $\hbar$ is the reduced
Planck constant~\cite{landau2013quantum}.   This leads to a total elastic cross
section $\sigma(E) = \sigma_0 (E/E_6)^{3/10} x_6^2$, where $\sigma_0 =
5/2\cdot3^{2/5}(1+\sqrt{5})\pi^{7/5}\Gamma(3/5)/(10\cdot 2^{3/5}) = 6.125
\cdots$.  We thermally average the loss rate coefficient by assuming that the
cold atoms  (typically with temperatures $\lesssim 1$~mK) are stationary
relative to the room temperature gas.  The result is
\begin{eqnarray}
	\langle K \rangle  & = & \frac{1}{\mathcal{Z}}\int d^3 p_h e^{-p_h^2/(2m_h k_B T)} K(E) \\
	& = &  \kappa  \left(\frac{\mu}{m_h} \frac{k_B T}{E_6}\right)^{3/10}  x_6^3 \frac{E_6}{\hbar} \propto \frac{(k_B T)^{3/10}}{m_h^{3/10}}C_6^{2/5},
\end{eqnarray}
where $\mathbf{p}_h$ is the initial momentum of the background gas molecule, $E =
(m_c/M)[p_h^2/(2 m_h)]$, $M = m_c+m_h$, $\kappa =
4\Gamma(9/5)\sigma_0/\sqrt{\pi} = 12.88\cdots$, and $\mathcal{Z}$ is the
partition function for the background gas.  In general, $E_6/k_B\approx 1$~mK
and $k_B T/E_6\gg 1$.  The last proportionality shows the dependence on $C_6$,
$m_h$, and $T$; surprisingly, it does not depend on $m_c$.

The largest correction to Eq.~\ref{eq:pressure} is the lack of a one-to-one
correspondence between a collision and the ejection of a cold atom from its
trap~\cite{Bali1999,Fagnan2009}.  To eject an atom, the final kinetic energy of
the initially cold atom must be at least $W$, the depth of a trap that is
equally deep in any direction.  Atoms are not ejected for scattering angles
$\theta_r$ less than the critical angle $\theta_c$, defined by
\begin{equation}
	\label{eq:critical_scattering_angle}
	\cos\theta_c = 1 - \frac{1}{2} \frac{m_c}{\mu}\frac{W}{E},
\end{equation}
as follows from energy and momentum conservation assuming a cold atom initially
at rest.  The loss rate coefficient for such glancing collisions with an
isotropic potential is
\begin{equation}
	\label{eq:glancing}
	K_{\rm gl}(W,E) = 2\pi \int_0^{\theta_c(W)} v\frac{d\sigma(E,\theta_r)}{d\Omega_r}\ \sin\theta_r d\theta_r,
\end{equation}
where $d\sigma/d\Omega_r$ is the differential cross section, where $\theta_c(W)$
is given by Eq.~\ref{eq:critical_scattering_angle}.  In the semiclassical
theory, the thermally-averaged result to first order in trap depth $W$ is
\begin{eqnarray}
	\label{eq:glancing_first_order}
 	\langle K_{\rm gl}(W)\rangle & = & \frac{1}{\mathcal{Z}}\int d^3 p_h e^{-p_h^2/(2m_h k_B T)} K_{\rm gl}(W,E) \\
	& \approx & \kappa \zeta \frac{m_c}{\mu}\frac{W}{E_6}\left(\frac{\mu}{m_h} \frac{k_B T}{E_6}\right)^{-1/10}x_6^3\frac{E_6}{\hbar},
\end{eqnarray}
where $\zeta = 25 \pi ^{13/10} [\Gamma (8/5)]^3/(4\cdot 6^{6/5} \sigma_0) =
0.3755\cdots$.  We find the higher order corrections numerically by integrating
\begin{equation}
	\label{eq:diff_cross_section}
	\frac{d\sigma}{d\Omega_r} = \left|\frac{1}{2 i (E/E_6)^{1/2}}\sum_{\ell=0}^\infty (2\ell+1)P_\ell(\cos\theta_r)\left(e^{2i\eta_\ell(E)}-1\right)\right|^2x_6^2,
\end{equation}
where $P_\ell(x)$ are the Legendre polynomials and $\eta_\ell(E)$ is given by Eq.~\ref{eq:phase_shift}.

\begin{figure}
	\center
	\includegraphics{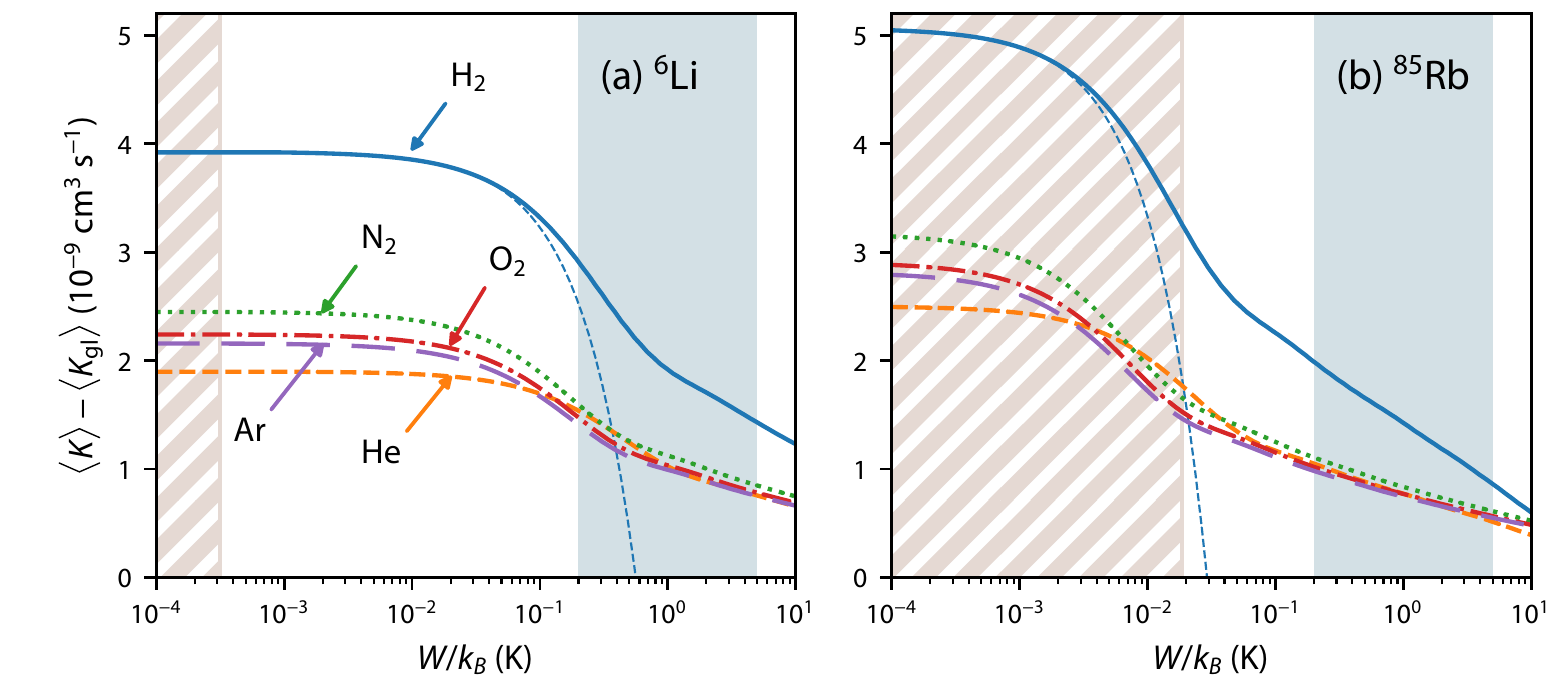}
    \caption{\label{fig:glancing}  (Color online) Glancing-collision-corrected loss
rate coefficient for ground-state $^{6}$Li($^2$S), panel (a), and
$^{85}$Rb($^2$S), panel (b), as a function of trap depth for various background
gases at $T=293$~K.   For H$_2$, the thin-dashed curve shows the first-order
result in $W$, Eq.~\ref{eq:glancing}. The red-striped (blue) shaded regions
highlight the accessible range of trap depths with a magnetic (magneto-optical)
trap.  Note that for magnetic trapping, we assume that cold atoms are in the
$F=I-1/2$ hyperfine state, which leads to different maximum trap depths for Li
and Rb.}
\end{figure}

These glancing collisions change the ideal CAVS operation (Eq.~\ref{eq:pressure}) to
\begin{equation}
	\label{eq:pressure_2}
	p = \frac{\Gamma}{\left<K\right>-\left<K_{\rm gl}(W)\right>} k_B T.
\end{equation}
Figure~\ref{fig:glancing} shows the CAVS loss rate coefficient with glancing
collisions, $\left<K\right>-\left<K_{\rm gl}(W)\right>$, for several cold atomic
species and room-temperature background gases as a function of trap depth based
on the numerical integration of Eq.~\ref{eq:diff_cross_section}.  This plot has
several interesting features.  First, for the same background gas, Rb, with its
larger van-der-Waals coefficients, has a larger loss rate coefficient than Li.
Second, $\langle K \rangle$ for H$_2$ collisions is twice as large as for other
gases, due primarily to its smaller mass.  Third, the first order behavior,
Eq.~\ref{eq:glancing}, is an excellent approximation until $[\langle K\rangle -
\langle K_{\rm gl}(W)\rangle]/\langle K\rangle\approx 0.9$.  At this point, the
linear behavior starts to give way to a logarithmic dependence on $W$. This
appears as a straight line on the log-linear scale.  In fact, $\langle K_{\rm
gl}(W)\rangle/\langle K\rangle\approx 0.1$ defines a crossover trap depth,
$W_c$, which scales as
\begin{equation}
W_c \propto \frac{m_h^{1/10}}{m_c^{1/2}(m_c+m_h)^{1/2}}\frac{1}{C_6^{3/10}}.
\end{equation}
Thus, for the same background gas, Rb, which is both more massive than Li and
has larger $C_6$ coefficients, has a smaller $W_c$.  As shown in
Fig.~\ref{fig:glancing}, the transition in the $\langle K\rangle-\langle K_{\rm
gl}(W)\rangle$ behavior occurs at higher $W$ for Li ($W_c/k_B \approx 0.5$~K)
compared to Rb ($W_c/k_B \approx 20$~mK).

There are two traps that are easy to realize in the p-CAVS given our design
constraints: a MOT and a quadrupole magnetic trap.  Each has a different trap
depth and, consequently, different fractions of glancing collisions.  MOTs
generally have depths ranging from 200~mK to 5~K depending on their parameters,
as shown in Fig.~\ref{fig:glancing}, where glancing collisions reduce the losses
by over one-half.  Quadrupole magnetic traps have depths of the order of 100~mK
or lower, determined by the atomic state.  As a result, the uncertainty budgets
associated with operating these two types of traps are  different.

The determination of $\Gamma$ from atoms contained within the traps is also
different.  In a MOT, the measurement proceeds by loading the trap and observing
the loss of atoms from the trap by continuously monitoring their fluorescence.
Thus, making a single MOT yields many points on the $N(t)$ curve.  This is in
contrast to operation with a quadrupole magnetic trap, which first requires
loading atoms into a MOT followed by optical pumping into the
magnetically-trapped atomic state.  After free evolution, the atoms in the
magnetic trap are recaptured into the MOT and counted by measuring the
fluorescence.   In this operation, a single load of the magnetic trap yields a
single point on the $N(t)$ curve.  Constructing a decay curve with a reasonable
signal to noise thus requires loading and measuring multiple times.  Thus, this
mode of operation is significantly slower than that of the MOT; however, as we
shall see, it is more accurate.

\subsection{Fast operation of p-CAVS: magneto-optical trap}
Operating the MOT as a pressure sensor presents several type-B (systematic)
uncertainties, some of which were anticipated in Ref.~\cite{Arpornthip2012}.
Glancing collisions are the dominant correction to the ideal CAVS operation in a
MOT.  Translating the loss rate of atoms from the MOT into a pressure therefore
requires knowledge of its trap depth.  Two trap-depth-measurement techniques
have been employed: inducing two-body loss with a known, final kinetic energy
with a catalyst laser~\cite{Hoffmann1996} and comparing the background-gas
induced MOT loss rates to a magnetic trap with known depth~\cite{VanDongen2011}.
These two methods have been shown to yield identical
results~\cite{VanDongen2011}.  Given their complexity, however, it is not clear
whether such measurements could be implemented in a sensor.

\begin{figure}
	\center
	\includegraphics{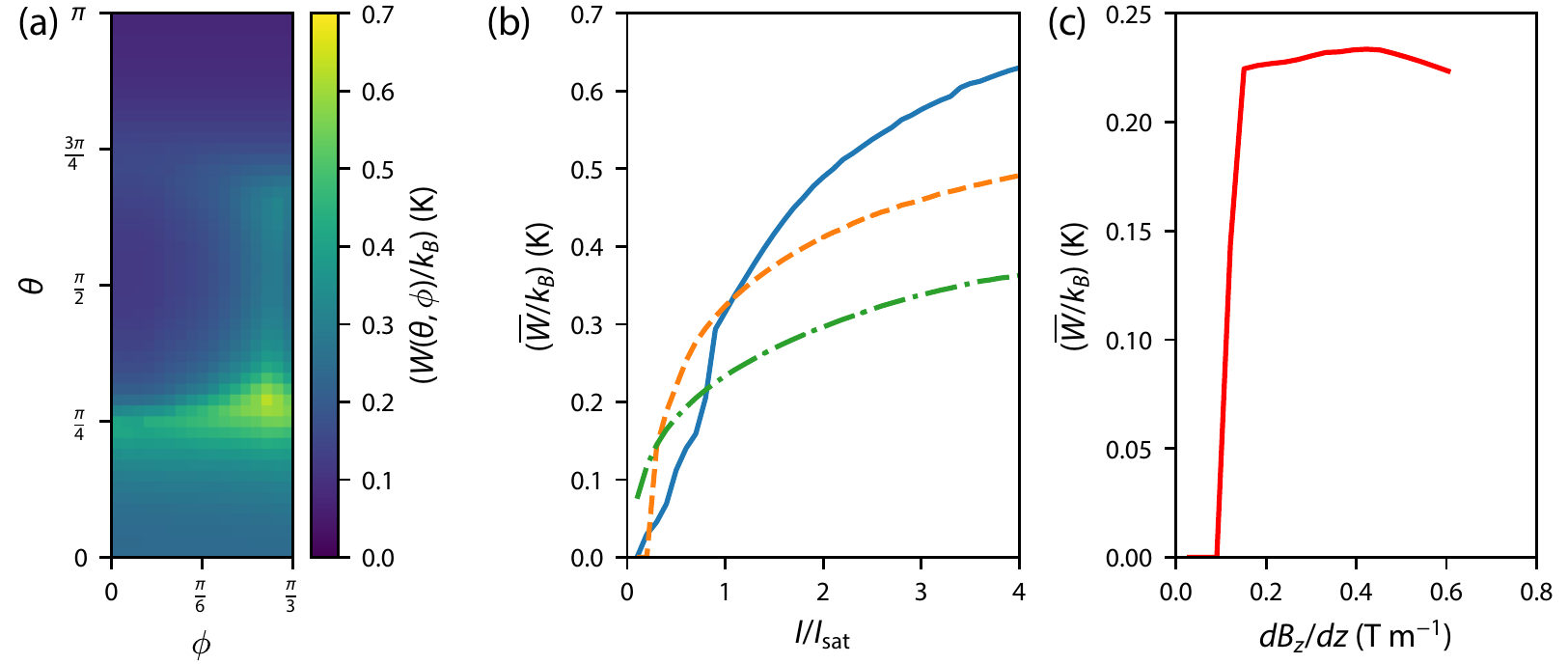}
    \caption{\label{fig:forces} Trap depth $W$ for a typical, three beam grating
    MOT for Li.  (a) Angularly-resolved $W(\theta,\phi)$ for an incident beam
    with  $I/I_{\rm sat} = 1$, $\Delta/\gamma = -1$, and
    $dB_z/dz=0.5$~T~m$^{-1}$.  (b) Average trap depth as a function of incident
    beam intensity for detunings $\Delta/\gamma = -3.0$ (solid blue), $-2.0$
    (dashed orange), and $-1.0$ (dashed-dot green) with
    $dB_z/dz=0.5$~T~m$^{-1}$.  (c) Average trap depth as a function of magnetic
    field gradient for $I/I_{\rm sat} = 1$ and $\Delta/\gamma = -1$.}
\end{figure}

Models of the trap depth of a MOT have been developed and find quantitative
agreement with measurements of two-body collisions between cold
atoms~\cite{Ritchie1994}.  The models  assume an atom with an optical cycling
transition between a ground state with electronic orbital angular momentum $L=0$
(S) and an excited state with $L=1$ (P).  (Here, we ignore effects due to
spin-orbit coupling and hyperfine structure.)  The non-conservative force on an
atom in a MOT results from the interplay of a spatially-varying magnetic field
$\mathbf{B}(\mathbf{r})$ and multiple laser beams $i$ with the same frequency
detuning $\Delta$ with respect to the atomic transition but different
wavevectors $\mathbf{k}_i$ and circular polarizations $\epsilon_i=\pm1$.  The
resulting force on the atom with position $\mathbf{r}$ and velocity $\mathbf{v}$
is
\begin{eqnarray}
	\mathbf{F}(\mathbf{r},\mathbf{v}) & = & \sum_{i}\sum_{m = -1}^1 P_i(m) \nonumber \\
	& & \times \frac{\hbar \mathbf{k}_i\Gamma}{2} \frac{s_i}{1 + \sum_j s_j + 4[\Delta - \mathbf{k}_i\cdot\mathbf{v} - (m \mu_B |\mathbf{B}(\mathbf{r})|/\hbar)]^2/\gamma^2}, \label{eq:higher_dim_force}
\end{eqnarray}
where $s_i=I_i/I_{\rm sat}$ is the saturation parameter of beam $i$ with
intensity $I_i$.  Here, the saturation intensity $I_{\rm sat} $ and linewidth
$\gamma$ are properties of the atom and $\mu_B$ is the Bohr magneton.  The
probability of making a transition to an excited angular momentum projection $m$
is
\begin{equation}
	\label{eq:trans_prob}
	P_i(m) = |d^1_{\epsilon_i m}(\pi/2-\xi_i)|^2 = \left\{
		\begin{array}{ll}
		\left(1-\epsilon_i\sin\xi_i\right)^2/4, &  m =\pm 1 \\
		(\cos^2\xi_i)/2, &  m = 0
		\end{array}
	 \right.,
\end{equation}
where $\xi_i$ is the angle between $\mathbf{k}_i$ and $\mathbf{B}(\mathbf{r})$
and $d^j_{mm'}(\theta)$ is a Wigner rotation matrix.

We model the MOT trap depth for the p-CAVS using
Eqs.~\ref{eq:higher_dim_force}--\ref{eq:trans_prob} with the beam geometries,
polarizations, and magnetic field specific for our device as shown in
Fig.~\ref{fig:basic_idea}b.  We use the magnetic field gradient
\begin{equation}
	\label{eq:mag_gradient}
	\mathbf{B}(\mathbf{r}) = \frac{dB_z}{dz}\left[z\hat{z} - \frac{1}{2} \rho\hat{\rho}\right]
\end{equation}
in cylindrical coordinates $\mathbf{r}= (\rho,\phi,z)$ with parameter $dB_z/dz$.
The magnetic field is zero at $\mathbf{r}=0$.  The diffraction grating shown is
positioned at $z_g=+5$~mm and is illuminated with a $\epsilon = +1$ polarized
Gaussian beam traveling along the $+\hat{z}$ direction.  The beam's $1/e^2$
radius is 15~mm.  The diffraction grating lines are made from superimposed
equilateral triangles.  The triangles continue outwards until clipped by a
circle with diameter 22~mm.  A central, triangle-shaped through-hole, fitting an
inscribed circle of radius 2.5~mm, produces a vacuum connection to the rest of
the chamber.  The three sides of the triangles form three grating sections that
each produce two beams with angle $\theta_d = \pi/4$ with respect to the normal
of the grating ($-\hat{z}$), one points toward the central axis of the MOT and
the other outwards.  Only the inward beams contribute to forming the MOT.  The
polarizations of these reflected beams is $\sigma^-$; their intensity profile is
assumed to be the same as the incident beam, but clipped according to the area
of the grating section and translated along its $\mathbf{k}_i$ vector.  The
grating produces no zero-order reflection and equal $\pm1$ diffraction orders
with efficiency $\eta=1/3$ and absorbs $1/3$ of the incident intensity.  The
resulting ratio of the reflected beam intensity to that of the incident is
$\eta/\cos\theta_d$, where the cosine describes the decrease in the beam's cross
section.

The magnetic field zero does not specify the center of the trap for a grating
MOT.  Unlike a standard 3D-MOT~\cite{FootBook} where $P_i(m=0)=0$ along
$\rho=0$, $P_i(m=0)$ is larger than $P_i(m=\pm1)$ for the beams reflected from
the grating, producing a position-independent force from these
beams~\cite{Vangeleyn2011}.  We find the trap center $\mathbf{r}_0 = (0,0,z_0)$
by placing an atom at rest at $\mathbf{r}=0$, integrating the equations of
motion (including the shape of the beams) and following its damped motion to the
center.   For alkali-metal atoms, MOTs are either overdamped or slightly
underdamped.  For our parameters, $z_0>0$.

The temperature of the cold-atom cloud is small compared to the trap depth;
therefore, the atoms are initially concentrated near the center of the trap.
After a collision with a background particle, they acquire momentum
$\mathbf{q}_c$ directed at azimuthal angle $\phi$ and polar angle $\theta$ in
the laboratory frame.  To determine the trap depth $W$, we can numerically
integrate the equations of motion starting from the center of the trap.  For
each pair of ($\theta,\phi$), the trap depth $W(\theta,\phi)$ is given by the
initial kinetic energy $q_e^2/(2m_c)$, where $v_{e} = q_e/m_c$ is the escape
velocity.

Figure~\ref{fig:forces}a shows $W(\theta,\phi)$ for a Li grating MOT with
$\Delta/\gamma=-1$, $dB_z/dz = 0.5$~T~m$^{-1}$, and the saturation parameter
$s=1$ for the incident beam.   We observe significant anisotropy in the trap
depth, varying from 0.1~K to 0.7~K  (only azimuthal angles of $0<\phi<\pi/3$ are
shown because of the three-fold symmetry of the grating MOT).  This is possible
because MOTs are overdamped: an atom launched from the center of the trap with
$q_c<q_e$ does not move chaotically through the trap, but instead quickly
returns to the center\footnote{This is in contrast to a conservative,
anisotropic magnetic trap, where an atom excited by a glancing collision will
chaotically orbit the trap center until it is ejected.}.  The polar angle at
which the trap depth is largest is $\theta=\pi/4$, corresponding an atom moving
directly into the reflected beams.  The azimuthal angle that maximizes the depth
is $\phi=\pi/3$, where two reflected beams both apply equal force.  Finally, the
shallowest direction corresponds to $\theta=\pi$, or into the incoming laser
beam.

The anisotropy of $W(\theta,\phi)$ complicates the calculation of $\langle
K_{\rm gl}(W)\rangle$.  The thermally averaged loss coefficient in this case
becomes
\begin{equation}
	\overline{\langle K_{\rm gl}(W)\rangle} = \frac{1}{\mathcal{Z}}\int d^3p_h e^{-p_h^2/(2m_h k_B T)} \int d\Omega_r\  v \frac{d\sigma}{d\Omega_r}\  H\left(W(\theta,\phi) - \frac{q_c^2}{2m_c}\right),
\end{equation}
where $H(x)$ is the Heaviside step function, $d\Omega_r =
\sin\theta_rd\theta_rd\phi_r$, and $\theta_r$ and $\phi_r$ are the scattering
angles.  Realizing that the angle between the initial $\mathbf{p}_h$ and final
$\mathbf{q}_c$ is uniquely determined by $\theta_r$, we interchange variables
and find
\begin{eqnarray}
	\overline{\langle K_{\rm gl}(W)\rangle} = \frac{1}{4\pi}\int d\Omega\  \langle K_{\rm gl}(W(\theta,\phi))\rangle,
\end{eqnarray}
where $d\Omega = \sin\theta d\theta d\phi$.   We compute an angle dependent
$\overline{\langle K_{\rm gl}(W)\rangle}$  using $W(\theta,\phi)$ and
Eq.~\ref{eq:diff_cross_section} for each $(\theta,\phi)$ and average over all
angles.   For the present work, we use the approximation $\overline{\langle
K_{\rm gl}(W)\rangle}\approx \langle K_{\rm
gl}(\overline{W(\theta,\phi)})\rangle$, where $\overline{W} = \int d\Omega\
W(\theta,\phi)/(4\pi)$, which is accurate within the currently known MOT
uncertainties (see below).

We have studied the angularly-averaged trap depth $\overline{W}$ for a Li
grating MOT to investigate the dependence on detuning $\Delta$, intensity of the
incident beam $I$, and magnetic field gradient.  The results are shown in
Fig.~\ref{fig:forces}.  As with a standard six-beam MOT, the trap depth
increases with increasing $s$ for a given $|\Delta/\gamma|$, shown in
Fig.~\ref{fig:forces}b.  For small $s$, the large $P_i(m=0)$ component of the
reflected beams creates a complicated dependence on $|\Delta/\gamma|$.  It also
causes a sudden breakdown of the trap for magnetic field gradients
$<0.1$~T~m$^{-1}$,  shown in Fig.~\ref{fig:forces}c.  This ``critical'' magnetic
field gradient is the gradient required to balance the force toward the grating
from the magnetic-field sensitive $m=+1$ component with the force away from the
grating from the magnetic-field insensitive $m=0$ component.

The uncertainty in the pressure due to uncertainty in the MOT's trap depth is
suppressed.  In particular, the fractional uncertainty in the measured pressure
is $\delta p/p= \delta \overline{W}/\overline{W}\log(\overline{W}/W_0)|$, based
on Eq.~\ref{eq:pressure_2} and $\langle K \rangle - \langle K_{\rm
gl}(\overline{W})\rangle \propto -A\log(\overline{W}/W_0)$ for MOTs, where $A$
and $W_0$ are constants that depend  on the background gas and sensor atom. For
Rb, $W_0/k_B\approx300$~K for most collisions other than H$_2$; for Li,
$W_0\approx 1000$~K for collisions other than H$_2$.   For example, consider an
uncertainty $\delta \overline{W}/\overline{W} \approx 20$~\% and
$\overline{W}/k_B\approx1$~K; here, $\delta p/p \approx 8$~\% for Rb and $7$~\%
for Li.   The actual uncertainty $\delta \overline{W}$ is currently difficult to
establish.  We have tested our model against the published data in
Ref.~\cite{VanDongen2011}, and find agreement to within the experimental error
bars for the smallest trap depths.  Based on this comparison, we currently
estimate the fractional uncertainty $\delta \overline{W}/\overline{W}$ of the
order of tens of per cent.  It is our intent to further improve the accuracy and
uncertainty of these models.

The second correction to the measured pressure by a MOT comes from the fact that
a non-negligible fraction of atoms are in the excited P state, which has
different $C_6$ coefficients compared to the ground S state (see
Tab.~\ref{tab:polarizability}).  With this correction, Eq.~\ref{eq:pressure_2}
becomes
\begin{equation}
	p = \frac{\Gamma}{(1-P_{\rm ex})\left< K-K_{\rm gl}(W) \right>_{\rm ground} +
	P_{\rm ex} \left<  K - K_{\rm gl}(W) \right>_{\rm excited}} k_B T,
\end{equation}
where $P_{\rm ex}$ is the probability of an atom to be in the excited state.
For grating MOTs, $\mu_B |B(\mathbf{r}_0)|/\hbar\ll\Delta$, and
\begin{equation}
	P_{\rm ex} = \frac{1}{2}\sum_{i}\frac{s_i}{1+\sum_{j} s_j +
	4(\Delta/\gamma)^2}\ .
\end{equation}
Typically, $s_i\approx 1$ and $\Delta/\gamma\approx -1$, making $P_{\rm ex}
\approx 25$~\%.  The uncertainty in $P_{\rm ex}$ is dominated by that of $s_j$,
which at best has $\delta s_j/s_j \approx 5$~\%, leading to $\delta P_{\rm
ex}/P_{\rm ex}\approx 12$~\%.  From our numerical results, $\left<K-K_{\rm gl}
(W)\right> \propto (C_6)^{0.35}$ in the MOT regime, and $\left<K-K_{\rm gl}(W)
\right>_{\rm excited}/\left< K-K_{\rm gl}(W) \right>_{\rm ground} \approx
(C_{6,{\rm P}}/C_{\rm 6,{\rm S}})^{0.35}$.  We estimate an uncertainty in the
ratio of 14~\% based on our uncertainty in $C_6$.   For a typical MOT, the
fractional uncertainty in the measured pressure is relatively small: $3$~\% for
both Li and Rb.  Note that in this analysis we neglect the possibility of
inelastic collisions with atoms in the excited state, which change the internal
state of the cold atom.  These effects will need to be further studied.

\begin{figure}
	\center
	\includegraphics{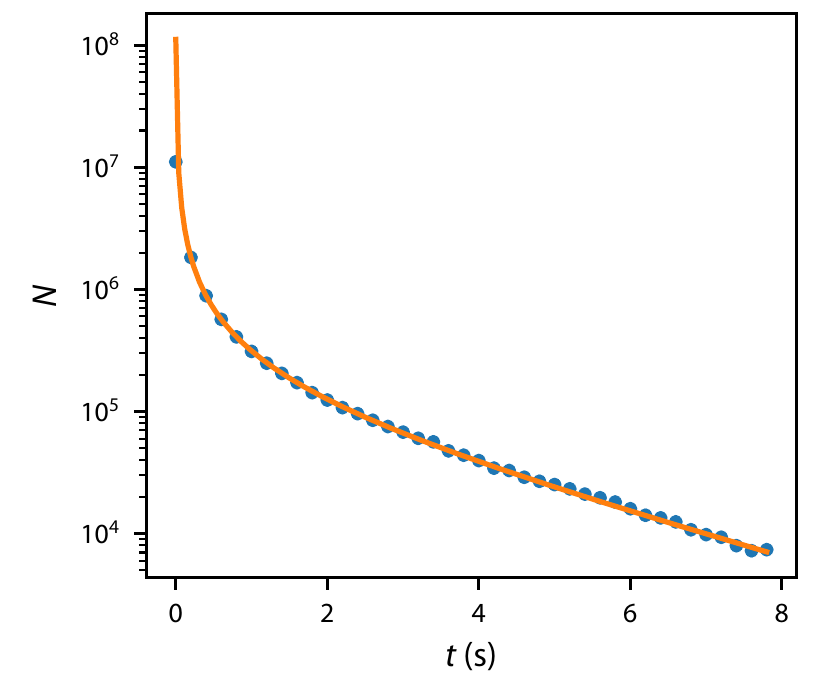}
    \caption{\label{fig:rejection} Number of $^7$Li atoms as a function of time,
    blue points, in a  standard six-beam MOT showing light-assisted two-body
    loss at early times ($t<1$~s) and exponential decay at longer times
    ($t>2$~s).  The orange curve shows a fit to Eq.~\ref{eq:two_body}.  The
    statistical uncertainty in the data is comparable to the size of the
    points.}
\end{figure}

Finally, another complication with using a MOT to measure pressure is the
presence of light-assisted collisions between cold
atoms~\cite{Gallagher1989,Sesko1989,Kawanaka1993,Browaeys2000}.  With these collisions, the number of
atoms in the trap $N$ obey
\begin{equation}
	\label{eq:two_body}
	\frac{dN}{dt} = - \Gamma N - K_2 N^2 - K_3 N^3 - \cdots,
\end{equation}
where $K_n$ is an $n$-body loss parameter that depends on the intensity and
detuning of the MOT light.  Figure~\ref{fig:rejection} shows such a decay curve
with large two-body loss measured in a standard, six-beam MOT of $^7$Li atoms.
The curvature observed at early times indicates the presence of two-body
collisions.  One can fit the data to Eq.~\ref{eq:two_body} to accurately
separate $n$-body loss from the exponential loss due to background gas
collisions.  No evidence of three- or higher-body loss was found in the data in
Fig.~\ref{fig:rejection}.  For these data, the MOT light is red-detuned to the
$F=2\rightarrow F'=3$ transition with $\Delta/\gamma = -2.0(1)$ and
$dB/dz\approx 0.5$~T/m.  Each of the six Gaussian beams has an intensity of
7.4(4)~mW/cm$^2$ with a $1/e^2$ diameter of 1.42(7)~cm.  Repump light is
provided by the $+1$ sideband of an electro-optic-modulator operating at
813~MHz.  Apporoximately 55~\% of the power remains in the carrier (red detuned
with respect to $F=2\rightarrow F'=3$) and $\approx 22$~\% of the power is in
the repump (tuned to resonance with the $F=1\rightarrow F'=2$ transition).

\subsection{Accurate operation: Quadrupole magnetic trap}
Unlike MOTs, magnetic traps are conservative traps: an atom's kinetic energy
must decrease by the same amount as its internal energy increases.  In free
space, Maxwell's equations only allow minima in $|\mathbf{B}(\mathbf{r})|$
(Earnshaw's theorem).   Therefore, only states whose internal energy
$\mathcal{E}$ increases with $|\mathbf{B}(\mathbf{r})|$, i.e.
$d\mathcal{E}/dB>0$, can be trapped.  In this section, we consider the
quadrupole trap generated by the MOT magnetic field given by
Eq.~\ref{eq:mag_gradient}.  This trap has its center at $\mathbf{r} = 0 \neq
\mathbf{r}_0$.

\begin{figure}
    \center
    \includegraphics{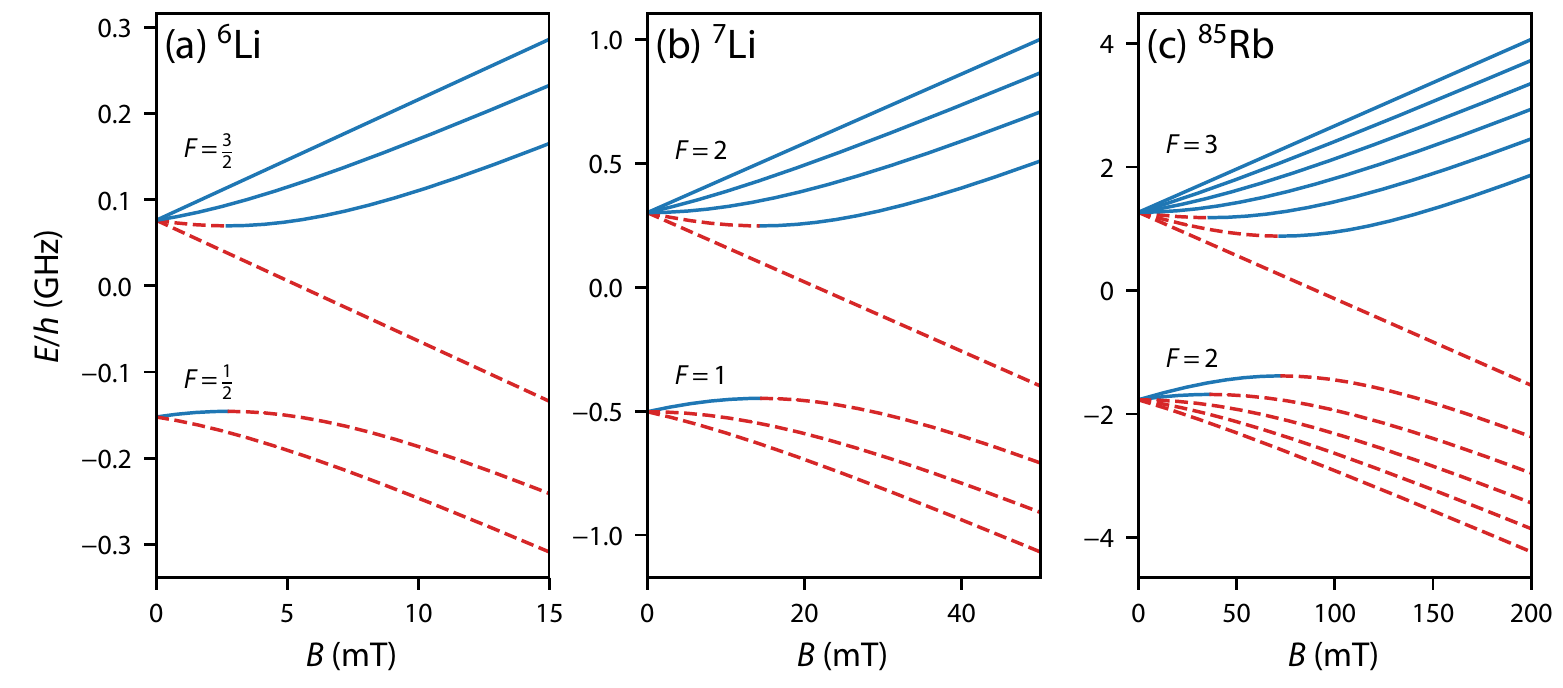}
    \caption{\label{fig:energy_levels} Energy of the magnetic sublevels as a
    function of magnetic field for (a) $^6$Li, (b) $^7$Li, and (c) $^{85}$Rb.
    Blue, solid curves (red, dashed) correspond to states that are (are not)
    magnetically trappable.  Note the different scales.}
\end{figure}


The energy of the internal states of $^6$Li($^2$S), $^7$Li($^2$S), and
$^{85}$Rb($^2$S) are shown in Fig.~\ref{fig:energy_levels}.  Here, we include
the hyperfine and Zeeman interactions.  The former gives rise to two
non-degenerate states at $\mathbf{B}=0$, denoted by $F=I\pm1/2$, where $I$ is
the nuclear spin.  For  $^6$Li, $^7$Li, and $^{85}$Rb, $I=1$, $3/2$, and $5/2$
respectively.  For non-zero $B$, the levels split according to projection
$m_F=-F,-F+1,\cdots,F$.

Magnetic traps in the limit $B\rightarrow\infty$ have infinite trap depth for
states with $F=I+1/2$ for these three atoms.  Hence, these states are
impractical for CAVS operation.  Instead, we focus on the state
$\left|F=I-1/2,m_F = -(I-1/2)\right>$, which has an energy
\begin{equation}
	\label{eq:breitrabi}
	\mathcal{E} = -\frac{\Delta_{\rm HF}}{2(2I+1)} + g_I m_F \mu_B B - \frac{\Delta_{\rm HF}}{2}\left[1 + \frac{4}{2 I+1}\frac{g m_F \mu_B B}{\Delta_{\rm HF}} + \left(\frac{g \mu_B B}{\Delta_{\rm HF}}\right)^2 \right]^{1/2}
\end{equation}
where $g=g_I-g_J$, $g_I$ and $g_J$ are the nuclear and electronic gyromagnetic
ratio respectively, and $\Delta_{\rm HF}$ is the zero-field energy splitting.
This state has a maximum energy at a finite $B_{\rm max}$ and trap depth $W_{\rm
max} = E(B_{\rm max}) - E(B=0)$.  Neglecting the $g_I m_F \mu_B B$ term in
Eq.~\ref{eq:breitrabi} yields
\begin{equation}
	B_{\rm max} \approx \frac{2I-1}{2I+1} \frac{\Delta_{\rm HF}}{g \mu_B}
\end{equation}
and
\begin{equation}
	W_{\rm max} \approx \Delta_{\rm HF}\left(\frac{1}{2}-\sqrt{\frac{2 I-1}{2(2I+1)^2}}\right).
\end{equation}
Table~\ref{tab:mag_trap_properties} lists $B_{\rm max}$ and $W_{\rm max}$ for Li
and Rb isotopes.  The uncertainty $B_{\rm max}$ and $W_{\rm max}$ is set by the
uncertainty in the atomic physics parameters, which are known to better than
1~ppm\footnote{Trap depths can be made arbitrarily smaller using a so-called RF
knife, which applies a radio-frequency magnetic field that couples a trapped
state to an untrapped state at a given magnetic field strength.  In this case,
the trap depth is set by the frequency of the oscillating magnetic field.}.

\begin{table}
	\center
	\begin{tabular}{l|c|c|c|c|}
	Species & $B_{\rm max}$ (mT) & $W_{\rm max}/k_B$ (mK) & $dB_z/dz$ (T m$^{-1}$) & $z_T$ (mm) \\
	\hline
	$^6$Li & 2.7168 & 0.31409 & 0.50 & 5 \\
	$^7$Li & 14.357 & 2.5946 & 0.50 & 30 \\
	$^{85}$Rb & 72.251 & 18.578 & 0.15 & 480 \\
	$^{87}$Rb & 244.30 & 62.971 & 0.15 & 1600 \\
	\hline\hline
	\end{tabular}
	\caption{\label{tab:mag_trap_properties} Energy-maximizing magnetic fields
	$B_{\rm max}$, resulting trap depths $W_{\rm max}$, typical magnetic field
	gradients used in a magneto-optical trap $dB_z/dz$, and resulting trap size
	$z_T =B_{\rm max}/(dB_z/dz)$ for various species.  Note that $B_{\rm max}$
	and $W_{\rm max}$ are typically known to within a ppm, while $dB_z/dz$ and
	$z_T$ are estimates.}
\end{table}

Using the $dB_z/dz$ for a MOT sets the characteristic size of the magnetic trap
through $z_T = B_{\rm max}/(dB_z/dz)$.  Table~\ref{tab:mag_trap_properties}
lists both $dB_z/dz$ and $z_T$.  The size of initial cold atom does not equal
$z_T$, but is set by its temperature out of the MOT, $\lesssim 1$~mK.  One then
expects from the virial theorem a cloud size $z_c \approx 5$~mm for Li and $z_c
\approx 20$~mm for Rb.  For $^6$Li, with $z_c>z_T$, this causes some loss of
atoms when transferred from the MOT to the magnetic trap.  For Rb, with
$z_c>z_g$, the cloud will expand into the grating, which is the closest
in-vacuum component.  This may require increasing the magnetic field gradient to
reduce the size of the initial cold-atom cloud.

The grating decreases the trap depth when $z_T>z_g$, as higher-energy atoms
eventually collide with and, most likely, stick to the grating.  (The classical
orbits in a quadrupole trap are not closed.)  The trap depth is then determined
by geometry, i.e., $W = |g m_F \mu_B (dB_z/dz) z_g|$; its fractional uncertainty
is set by $\delta z_g/z_g$ and $\delta(dB_z/dz)/(dB_z/dz)$.  For Rb with
$z_g=5(1)$~mm and $dB_z/dz=0.15(2)$~T~m$^{-1}$, $W=1$~mK and $\delta W/W \approx
25$~\%.  In a magnetic trap, Eq.~\ref{eq:glancing_first_order} is an excellent
approximation and thus the fractional uncertainty in the glancing collision
fraction is also 25~\%.

Glancing collisions in a magnetic trap can still lead to loss of atoms from the
trap\footnote{This is in contrast to a MOT, which recools atoms not ejected from
the trap.}.  The average energy deposited by a glancing collision is $Q=W/2$.
Moreover, the average amount of energy necessary to cause ejection is $\approx
W-k_BT_c$, where $T_c$ is the temperature of the cold atoms.  Consequently,
starting in the limit where $k_B T_c\ll W$, glancing collisions only heat the
gas and the loss rate is given by $\Gamma = n(\langle K\rangle - \langle K_{\rm
gl}(W)\rangle)$.  As the trapped gas warms and $k_B T_c\gtrsim W/2$, more of the
glancing collisions start contributing to the loss and $\Gamma$ approaches $n
\langle K\rangle$.  Because $\Gamma$ depends on $T_c$ and time, we expect that
this will cause non-exponential decay and thus may be separable in a manner
similar to the $n$-body loss of Eq.~\ref{eq:two_body}.  This heating through
glancing collisions is a problem that we also anticipate with the
laboratory-scale CAVS and are currently performing Monte-Carlo studies to
understand.   For the present analysis, however, we take the measured pressure
with these glancing collisions to be the mean of the two limits,
\begin{equation}
	p = \frac{\Gamma}{\left< K\right> - \left<K_{\rm gl}(W)\right>/2} k_B T,
\end{equation}
with a fractional uncertainty $\delta p/p\approx \langle K_{\rm
gl}(W)\rangle/(2\langle K \rangle$).

Majorana spin-flip losses also contribute to the loss in a quadrupole trap
\footnote{The laboratory-scale CAVS uses a Ioffe-Pritchard magnetic trap to
suppress Majorana loss.}.  Because the trap has a location where $B=0$, atoms
that pass sufficiently close to the center can undergo a diabatic transition
into the untrapped spin state.  Reference~\cite{Petrich1995} estimates the decay
rate to be
\begin{equation}
	\Gamma_{\rm Majorana} \approx \frac{\hbar}{m_c z_c^2}\ .
\end{equation}
This estimate was found to be about a factor of 5 too small for the experimental
data in Ref.~\cite{Petrich1995}.  For $^7$Li, $\hbar/m_c \approx
9\times10^{-3}$~mm$^2$~s$^{-1}$ and $\Gamma_{\rm Majorana}\approx
10^{-3}$~s$^{-1}$; for $^{85}$Rb, $\hbar/m_c \approx
7\times10^{-4}$~mm$^2$~s$^{-1}$ and $\Gamma_{\rm Majorana}\approx
10^{-5}$~s$^{-1}$.  These loss rates could be mistaken as N$_2$ pressures of
approximately $10^{-9}$~Pa and $10^{-11}$~Pa, respectively.  It is, however,
possible that the Majorana loss is not exponential and could be separated out by
fitting, much like with two body loss in a MOT.

\subsection{Summary of uncertainties}
\label{sec:princ_of_operation_uncertainties}
\begin{table}
	\center
	\begin{tabular}{l|cc|ccc|}
    	 & \multicolumn{2}{c|}{{\bf MOT (fast)}} & \multicolumn{3}{c|}{{\bf Magnetic trap (slow)}} \\
		{\bf Effect}	& Li & Rb & $^6$Li & $^7$Li & $^{85}$Rb \\
        \hline
        Glancing collisions & $7$~\% & $8$~\% & $10^{-4}$ & $10^{-3}$ & 2~\% \\
        Excited state fraction & 3~\% & 3~\% & \multicolumn{3}{c|}{n/a} \\
        Majoranna losses & \multicolumn{2}{c|}{n/a} & 5~\% & 5~\% & 0.05~\% \\
        Loss rate coefficient & 5~\% & 5~\% & 5~\% & 5~\% & 5~\% \\
        \hline\hline
        Total & 9~\% & 10~\% & 7~\% & 7~\% & 5.5~\% \\
        \hline\hline
    \end{tabular}
    \caption{\label{tab:err_summary} Estimated uncertainty in the pressure from
    various effects associated with the p-CAVS operating at $10^{-7}$~Pa using a
    magneto-optical trap (MOT, left) and quadrupole magnetic trap (right).  Note
    that loss rate coefficient here refers to the ground-state loss rate
    coefficient.  Totals are quadrature sums.  See text for details.}
\end{table}
Table~\ref{tab:err_summary} shows the estimated type-B uncertainties in a p-CAVS
device.  The uncertainties are roughly equal for Li and Rb.
Table~\ref{tab:err_summary} does not include any uncertainties due to the
background gas composition; the composition is assumed to be known. Additional
requirements for a vacuum gauge, explored in the next section, therefore will
dictate our choice of sensor atom.

While we have focused thusfar on type-B uncertainties, it is important to note
there are type-A uncertainties as well.  In particular, we anticipate the
dominant type-A uncertainty to be statistical noise in the atom counting.  The
fit shown in Fig.~\ref{fig:rejection} has a relative uncertainty $\lesssim 1$~\%
with approximately 10~s of data.  Translated into a pressure sensitivity
(assuming N$_2$ as the background gas, $W=0$, and room temperature), this
corresponds to $\approx 10^{-8}$~Pa/$\sqrt{\mbox{Hz}}$.

\section{Details of the planned device}
\label{sec:field_operation}
In addition to the Quantum-SI requirements of being primary and having
uncertainties that are fit for purpose, a deployable vacuum gauge should satisfy
the following requirements:
\begin{enumerate}
	\item It must be able to withstand heating, in vacuum, to temperatures
approaching 150 C to remove water from the  surfaces and minimize outgassing of
the metal components.  After such a heat treatment, the predominant outgassing
component will be hydrogen gas trapped within the bulk of the stainless steel,
which can only be removed by heat treatment at temperatures exceeding 400 C.
    \item It must not affect the background gas pressure it is attempting to
measure, or the extent to which it does must be quantified and treated as a
type-B uncertainty.
    \item It must minimize its long-term impact on the vacuum chamber to which
it is coupled.
\end{enumerate}
The design shown in Fig.~\ref{fig:basic_idea} incorporates these additional
requirements, as detailed below.

\subsection{Sensor atom}
By far, the most commonly laser cooled atomic species is Rb, which offers easily
accessible wavelengths for diode lasers and easy production inside vacuum
chambers.  As a result, much work has focused on miniaturizing Rb-based cold
atom technology.  On the other hand, Rb has a high saturated vapor pressure of
$2 \times 10^{-5}$~Pa~\cite{Alcock1984} at room temperature, which threatens to
contaminate the vacuum it is attempting to measure.  Second, Rb precludes baking
a vacuum chamber, because its vapor pressure of $3\times 10^{-1}$~Pa at
150~$^\circ$C may cause any small, open source of Rb to be depleted during a
bake.

Lithium, on the other hand, has a saturated vapor pressure of
$10^{-17}$~Pa~\cite{CRC2016} at room temperature, the lowest of all the
alkali-metal atoms.  This limits its contamination of the vacuum chamber.  At
150~$^\circ$C, the saturated vapor pressure is approximately $10^{-9}$~Pa, low
enough to allow the vacuum chamber to be baked.

\subsection{The trap}
The magneto-optical trap itself is a novel design, and its features and
performance will be detailed elsewhere.  In short, a collimated,
circular-polarized beam reflects from a nanofabricated triangular diffraction
grating to produce three additional inward-going beams, the minimum needed for
trapping. To generate the quadrupole magnetic field for the MOT, we
intend to use neodymium rare-earth magnets mounted {\it ex-vacuo}. They are
removable during baking, so as to not change their remnant magnetization.

An aperture in the chip allows light and atoms to pass through the chip.  The
source is positioned behind the chip and the thermal atoms are directed toward
the aperture.  Light passing through the aperture can slow the atoms emerging
from the source.   We tailor the magnetic field profile along the vertical axis
such that it starts linearly near the center of the MOT and smoothly transforms
into a $\sqrt{z}$ behavior near the atomic source.  This creates an integrated
Zeeman slower that enhances the loading rate of the MOT.   Finally, the aperture
acts as a differential pumping tube, limiting the flow of gas from the source
region to the trapping region of the device.

\subsection{Beam shaping and detection}
Laser light is delivered into the p-CAVS using a polarization-maintaining
optical fiber with a lens for collimation and a quarter-waveplate for generating
circular polarization.  These components are maintained {\it ex-vacuo} and can
be removed during installation to prevent breakage and baking to prevent
misalignment.  The light travels through a fused-silica viewport on the top of
the vacuum portion of the device.

Detection of the atoms can be accomplished through the same viewport, using a
beamsplitting cube to separate the incoming light from the fluorescence light
returning from the atoms in the  MOT.  An apertured photodiode (not shown) with
an appropriate imaging lens will be used to detect the fluorescence.

\subsection{Atomic Source}
One problem that must be overcome with Li is building a thermal source
that is  UHV or XHV compatible.  Heating the source to the necessary 350
$^\circ$C to produce Li vapor while maintaining a low outgassing rate is a
challenge.

We recently demonstrated a low-outgassing alkali-metal dispenser made from
3D-printed titanium~\cite{Norrgard2018}.  The measured outgassing level,
$5(2)\times10^{-7}$~Pa~l~s$^{-1}$, would establish the low-pressure limit of the
gauge.  For example, an effective pumping speed\footnote{The effective pumping
speed is determined by the combination of pumping speed and conductance of the
components leading to the pumps.} of 25~l/s between the pCAVS and the chamber to
which it is attached will produce a constant pressure offset of approximately
$10^{-8}$~Pa relative to the pressure in the chamber under test.  One can
decrease this offset by adding pumps to the source portion of the pCAVS.  As
currently envisioned, the titanium dispenser will be surrounded by a
non-evaporable getter pump, created by depositing a thin layer of Ti-Zr-V onto a
formed piece of metal.  Assuming roughly 100~cm$^2$ of active area, this
translates to an approximate pumping speed of 100~L/s~\cite{Erjavec2011} with a
capacity of the order of $0.1$~Pa~l~\cite{Bansod2012}.  Such a pump will reduce
the pressure offset to $10^{-11}$~Pa and have an estimated lifetime of $10^8$~s,
comparable to the lifetime of the dispenser.  Further improvements can be made
by minimizing the creation of other lithium compounds when loading the lithium
into the dispenser~\cite{Norrgard2018}.

For the p-CAVS to be accurate, the flow of alkali-metal atoms must be turned off
while measuring the lifetime of the cold atoms in the trap.  Otherwise,
collisions between hot atoms from the source and cold, trapped atoms will cause
unwanted ejections.  These collisions have a loss rate coefficient that is
almost an order of magnitude larger than those due to other gasses.  To stop the
flow of atoms, our current design incorporates a mechanical shutter.

We are also considering other more speculative sources of lithium. Lithium, like
other alkali-metal atoms, can be desorbed from surfaces using UV
light~\cite{Barker2018}.  However, UV light also desorbs other, unwanted species
from surfaces, such as water and oxygen~\cite{Halama1991, Herbeaux1999,
Koebley2012}, increasing their background gas pressures.  In a recent
experiment~\cite{Barker2018}, we observed that the increase in pressure due to
unwanted gasses is significantly smaller than our low-outgasssing lithium
dispensor.  In addition, light-assisted desorption should be nearly
instantaneous with application of the light, eliminating the need for a
mechanical shutter.  The combination of low-outgassing and instantaneous
response make light assisted desportion an attractive source for the p-CAVS.
Finally, a source based on electrically-controlled chemical reactions, like
those in a battery, may also work as a nearly instantaneous source of lithium
with low outgassing~\cite{Kang2017}.

\section{Conclusion}
\label{sec:conclusion}
Our group is currently in the process of building a portable cold-atom vacuum
standard, the p-CAVS.  This gauge will be based on recent advances in grating
MOT technology and fit in a footprint equal to that of commonly used gauges for
this vacuum regime like Bayard-Alpert ionization and extractor gauges.  As part
of the emerging Quantum-SI paradigm, our device is primary (traceable to the
second and the kelvin) and has errors that are well-characterized and fit for
purpose.

There are two atom traps that we can operate with this gauge, each offering
different performance but also different speed.  The estimated uncertainties
discussed in the previous sections are summarized in Tab.~\ref{tab:err_summary}.
We find that the pressure uncertainty from the MOT is only slightly worse than
the magnetic trap.  These estimates, however, depend on the accuracy of the
semiclassical model of $\langle K \rangle$ and $\langle K_{\rm gl}(W)\rangle$
and are subject to change.  In a parallel effort, we are constructing a
laboratory-scale standard in which we intend to measure both $\langle K \rangle$
and $\langle K_{\rm gl}(W)\rangle$ to better than 5~\% accuracy.

\appendix
\section{Atom trapping: a short introduction}
\label{app:atom}

\begin{figure}
	\center
	\includegraphics{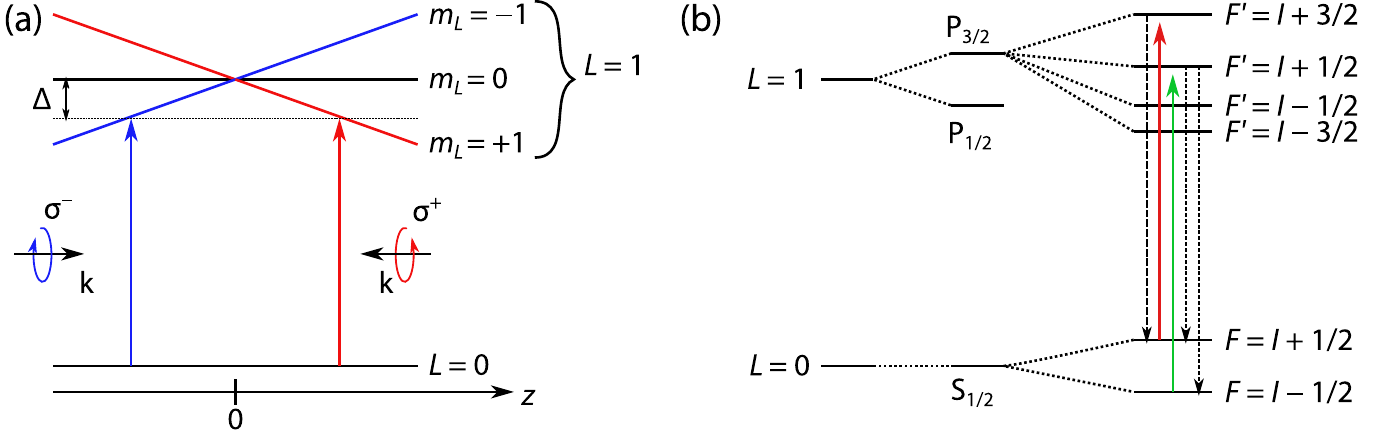}
	\caption{\label{fig:tutorial} (a) Schematic for a one-dimensional MOT.  Two
	beams with opposite circular polarizations (measured along $\hat{z}$) and
	zero-field detuning $\Delta$ are incident upon atoms in a magnetic field
	gradient (i.e, Eq.~\ref{eq:mag_gradient}).  The field is zero at $z=0$.
	This gradient splits the magnetic sublevels of the upper orbital angular
	momentum state into three.  (b) Hierarchy of splittings of a realistic
	alkali-metal atom. The orbital angular momentum states $L=0$ (S) and $L=1$
	(P) used in (a) are first split into states denoted by $L_J$ by spin-orbit
	interactions with total electronic angular momentum $J$.  These levels are
	again split when the nuclear spin $I$ is coupled in via the hyperfine
	interaction to $J$, creating states of total atomic angular momentum $F$.
	One typically operates the MOT on the $F=I+1/2$ to $F'=I+3/2$ transition
	(red arrow); however, because of off-resonant transitions between $F=I+1/2$
	to $F'=I+1/2$, a ``repump'' laser is added (green arrow).  The dashed arrows
	show possible decay channels from excited states to the ground state
	manifold by spontaneous emission.}
\end{figure}

Here, we provide a brief explanation of magnetic-optical trapping and magnetic
trapping, with a particular focus on the loading of atoms from one to the other.
For a more thorough introduction, the interested reader can consult
Refs.~\cite{FootBook,MetcalfBook}.

MOTs cool and trap atoms by a combination of the Doppler effect and spatially
varying light forces.  The forces arise from light pressure: when an atom
scatters a photon from a laser with wavevector $\mathbf{k}$, it receives a
momentum kick $\hbar \mathbf{k}$. The characteristic timescale for this process
is the excited state lifetime $1/\gamma$.

The typical MOT is depicted in Fig.~\ref{fig:tutorial}a in one dimension for an
atom with electronic orbital angular momentum $L=0$ in the ground state and
$L'=1$ in the excited state and projections $m_L$ of that angular momentum along
this direction.  First, consider an atom at some distance $+z$ with zero
velocity.  With the appropriately chosen polarizations, the right- (left-) going
beam couples the $m_L=0$ to $m_L'=+1$ ($m_L=0$ to $m_L'=-1$), as indicated by
the colors.  The Zeeman effect due to the magnetic field gradient shifts the
$m_L=0$ to $m_L'=+1$ transition into resonance with the leftward going laser,
while the rightward-going laser is shifted out of resonance with $m_L=0$ to
$m_L'=-1$ transition.  This causes the atom to scatter photons from the leftward
going beam and be pushed back toward the origin.  The two laser beams
interchange their roles for an atom placed at $-z$, causing the atom to be again
pushed toward the origin.  Second, consider the center of the trap where the
magnetic field is zero and the $m'_L$ levels are degenerate.
(Figure~\ref{fig:tutorial}a depicts a stationary atom.)  If the atom is moving
with velocity $+v$ ($-v$), the Doppler effect will shift the left (right) moving
beam into resonance and the atom will scatter photons and be slowed.  This is
the slowing or cooling force of a MOT.

This picture is further complicated by the presence of additional angular
momentum states in the atom, as shown in Fig.~\ref{fig:tutorial}b.  All
alkali-metal-atom MOTs operate on an electron orbital angular momentum $L=0$ (S)
to $L=1$ (P) transition.  However, the atom also has an electron spin $S=1/2$,
and the total electronic angular momentum is $\mathbf{J}=\mathbf{L}+\mathbf{S}$.
This results in a single ground state with $J=1/2$ and two excited states with
$J'=1/2$ and $J'=3/2$.  The degeneracy of the two excited states is broken by
spin-orbit coupling.  This presents us a choice of whether to operate a MOT on
the P$_{1/2}$ state (the D1 line) or P$_{3/2}$ state (the D2 line).  In general,
one wants the transitions driven in laser cooling to be ``cycling'' transitions:
the excited state only decays back to the original ground state.  This condition
is most easily achieved on the    $J=1/2$ to $J'=3/2$ transition and, therefore,
most MOTs operate on the D2 line.

This picture must also include the nuclear spin, which adds to $J$ to make a
total angular momentum $\mathbf{F}=\mathbf{I}+\mathbf{J}$.  For the ground state
with $J=1/2$, this makes two states $F=I\pm 1/2$ (for $I>1/2$) that are split by
the hyperfine interaction.  For the excited $J'=3/2$, it creates four states.
The cycling transition is once again found on the $F = I+1/2$ to $F' = I+3/2$
transition, which can only decay back to $F = I+1/2$ (see the dashed decay paths
in Fig.~\ref{fig:tutorial}b).

The hyperfine splitting in the excited state, however, is not sufficiently large
compared to the excited state lifetime to completely prevent transitions between
$F = I+1/2$ to $F' = I+1/2$.  If an atom is driven to this excited state, it can
decay by spontaneous emission into either of the $F=I\pm 1/2$ ground states.
Typically, as depicted in Fig.~\ref{fig:tutorial}b, one must apply a second
laser to ``repump'' the atoms out from $F=I-1/2$ back to $F=I+1/2$.

The repump laser can also be used to transfer atoms into a magnetic trap in a
simple way.  By merely turning off the repump laser, all atoms will eventually
find themselves in the $F=I-1/2$ ground state.  After this occurs, all lasers
can be turned off and the atoms that happened to be pumped into the
$m_F=-(I-1/2)$ state are magnetically trapped.  This is the simplest means to
load a magnetic trap from a MOT.  By re-applying both lasers, the atoms trapped
in the magnetic trap can be brought back into the MOT and counted.

\section*{References}
\bibliography{main}

\end{document}